# Daily Groundwater Monitoring Using Vehicle-DAS Elastic Full-waveform Inversion

Haipeng Li[1]*, Jingxiao Liu[2], Shujuan Mao[3], Siyuan Yuan[1], Robert G. Clapp[1], and Biondo L. Biondi[1]

[1]Department of Geophysics, Stanford University, Stanford, 94305, CA, USA.
[2]Senseable City Laboratory, Massachusetts Institute of Technology, Cambridge, 02139, MA, USA.
[3]Department of Earth and Planetary Sciences, Jackson School of Geosciences, The University of Texas at Austin, Austin, 78712, TX, USA.

*Corresponding author. E-mail: haipeng@sep.stanford.edu

## Abstract

Understanding groundwater dynamics is critical for sustainable water management, particularly as climate extremes intensify. However, the resolutions of existing subsurface observational tools are still inadequate for detailed aquifer monitoring and imaging. We introduce an innovative technique for groundwater monitoring using time-lapse full-waveform inversion, leveraging fiber-optic cables as seismic sensors and vehicular traffic as repetitive seismic sources. Over a two-year period along Sandhill Road, California, this approach captures detailed spatiotemporal S-wave velocity variations, revealing a 2.9% reduction corresponding to a 9.0-meter groundwater table rise after atmospheric-river storms in Water Year 2023. Notably, this approach enables the high-resolution daily analysis of rapid aquifer responses. We observe spatially inhomogeneous velocity changes, with less reduction beneath impervious paved zones than under grassy areas, underscoring the impact of urbanization on the natural recharge of aquifers. Our findings highlight the potential of Vehicle-DAS FWI for high-resolution daily monitoring and quantitative spatiotemporal characterizations of groundwater systems.

## Introduction

Groundwater is an overarching component of global freshwater resources, contributing over 60% of California's water supply during droughts (*1*). It supports drinking water supplies, agriculture, industry, and ecosystems, especially in regions lacking surficial water reservoirs. In recent years, historical droughts have accelerated groundwater depletion globally (*2*), while extreme precipitation events, such as California's 2023 atmospheric rivers, have led to rapid replenishment of aquifers (*3*). Despite the vital role of groundwater, its spatiotemporal variability remains poorly understood due to deficient observational resolutions to characterize the highly heterogeneous subsurface structures and dynamics. Monitoring groundwater at daily to weekly timescales across complex hydrogeological settings is essential for optimizing recharge and pumping strategies, facilitating sustainable water management, and enhancing the climate resilience of the water supply system (*4*).

Aquifer monitoring has benefited from various successful techniques, though some challenges remain to be addressed. Groundwater monitoring wells provide precise in-situ measurements of hydraulic head but lack sufficient spatial coverage (*5*). In contrast, satellite gravimetry enables large-scale monitoring of terrestrial water storage (*6*, *7*), and GPS- and InSAR-derived ground deformation data deliver indirect water level estimates over broad areas (*8*-*12*). Time-lapse geophysical approaches offer high-resolution insights into groundwater dynamics, including airborne electromagnetic, ground-penetrating radar, and active/passive source seismic methods (*13*-*16*). Seismic monitoring can detect subtle velocity changes associated with variations in pore pressure, saturation, porosity, and temperature (*17*, *18*), providing valuable proxies for groundwater changes (*1*, *19*, *20*). For instance, Mao et al. (*1*) showcased the use of coda waves from passive seismic interferometry to monitor basin-scale groundwater systems. Despite these advancements, existing methods face trade-offs in spatial and temporal resolutions, depth sensitivity, cost, and logistical complexity, underscoring the need for scalable, high-resolution, and cost-effective solutions for aquifer monitoring.

Distributed Acoustic Sensing (DAS) has emerged as a promising technology for seismic monitoring. By transforming existing telecommunication fiber-optic cables into dense sensor arrays, DAS offers both scalability and affordability. It enables meter-scale data acquisition over kilometers of distance (*21*, *22*), overcoming the limitations of sparsely deployed seismometers and facilitating high-resolution subsurface imaging (*18*, *23*-*25*). Tracking temporal variations, however, often entails daily active-source seismic surveys, which remain costly even with the use of dense DAS sensors. Daily vehicular traffic provides a passive, consistent seismic surface wave source for cost-efficient subsurface monitoring (*26*-*29*). In this study, we adopt a recently developed targeted interferometry approach, which facilitates repeatable retrieval of time-lapse Virtual Source Gathers (VSGs) while avoiding the requirement of uniformly distributed noise sources in ambient noise interferometry methods (*30*).

Monitoring with high spatiotemporal is essential in urban settings, where groundwater fluctuations can vary considerably over short distances and timescales, particularly during extreme hydroclimate events like atmospheric river storms or severe droughts. To achieve tens-of-meter resolution in space and daily resolution in time, we develop time-lapse elastic Full-Waveform Inversion (FWI) of VSGs from targeted interferometry. FWI method directly inverts full seismic waveforms and has shown superior resolution and sensitivity to subsurface fluid monitoring compared to conventional tomographic approaches (*31*-*37*). Surface-wave FWI, in particular, excels in resolving near-surface structures by accounting for complex surface wave propagation, compared to traditional dispersion curve analysis (*38*-*40*). Furthermore, unlike conventional seismic coda-wave monitoring that yields relative, spatially averaged velocity changes, surface-wave FWI can locate the absolute subsurface velocity variations at the sub-wavelength resolution, enabling a refined understanding of groundwater systems.

This study applies daily time-lapse elastic FWI to Vehicle-DAS data collected over two years (Dec. 2021 – Mar. 2024) along a 1 km section of Sandhill Road, California, as part of the Stanford DAS-2 urban experiment (see Fig. 1a). Inverted spatiotemporal variations in S-wave velocity ($v_s$) reveal groundwater dynamics driven by seasonal and meteorological forces. Velocity reductions of up to 2.9% correspond to heavy precipitation in Water Year 2023. Notably, FWI results reveal that an Extreme Precipitation Event (EPE) on December 31, 2022, resulted in a rapid, significant groundwater table rise. The observed velocity changes likely result from pore-pressure variations, as further supported by poroelastic simulations. These findings demonstrate that daily FWI using Vehicle-DAS data enables effective, high-resolution groundwater monitoring, providing a scalable and economical solution for characterizing aquifers in space and time. Furthermore, this capability could be expanded to monitor 3-D aquifers systems by interrogating telecommunication fiber networks with 2-D geometries.

## Results

### Groundwater monitoring along Sandhill Road, California

Our study area features a diverse urban landscape, including both impervious paved surfaces (at a parking lot) and highly pervious areas (at grassy lawns), as shown in Fig. 1b. The site is situated primarily on late Pleistocene alluvial fan deposits and early to mid-Pleistocene stream terrace deposits, according to a previous study of spectral-analysis-of-surface-waves (SASW) survey (see Fig. 1b) (*41*).

Our continuous DAS dataset, spanning over two years, provides a unique opportunity to investigate groundwater dynamics in response to both the annual hydrological cycle and sub-seasonal rainfall events. California experienced record-dry conditions in Water Year 2022, including the driest January to October on record (*42*). The Water Year 2023, however, brought the longest sequence of consecutive atmospheric rivers in 70 years, with above-average precipitation lifting the state out of drought conditions (*42*). We use local precipitation and groundwater level measurements from a nearby meteorological station (code KCAMENLO88, see sky-blue triangle in Fig. 1a) and groundwater monitoring well (code 06S03W02D032, see orange circle in Fig. 1a) to validate our seismic observations.

To achieve highly repeatable time-lapse seismic imaging, we utilize traffic-generated seismic signals recorded by DAS arrays for seismic interferometry (see Materials and Methods: Time-lapse targeted interferometry). In urban environments, ambient seismic fields are often affected unevenly distributed noise sources and exhibit considerable spatiotemporal variations (*43*). In contrast, traffic-induced signals serve as highly repetitive sources that generate broadband surface waves (*30*). Fig. 1c-e display the preprocessed DAS data, vehicle-induced surface wave energy (>1 Hz), and the retrieved VSG for a source at 610 m along the fiber on December 3, 2022. The obtained VSG is predominantly composed of surface waves, whereas P-wave energy is less pronounced.

## Near-surface Structure Variability

Baseline FWI was performed using a multi-scale inversion strategy to resolve near-surface structures (see Materials and Methods: Time-lapse elastic full-waveform inversion). The inverted baseline $v_s$ model reveals lateral and vertical variability of the near-surface structures, showing consistency with the surface conditions (see Fig. 2a and b). In the upper 5 m, we observe a low-velocity layer with $v_s$ around 300 m/s, characteristic of soft and unconsolidated sediments. A distinct low-velocity zone at shallow depth appearing from 0.65 to 0.75 km in horizontal distance coincides with a grassy area, likely reflecting the rich moisture in the loose soil beneath. In contrast, we observe higher velocity anomalies with $v_s$ of 500 to 600 m/s between 0.45 and 0.65 km in distance, which are beneath a parking lot at the ground surface with concrete pavement. The high-velocity structure observed between 0.70 and 0.85 km (green-reddish area), which extends to greater depths, is likely associated with the local Stanford fault (see fault line in Fig. 1).

At greater depths, a layered structure is revealed. A high-velocity layer with $v_s$ of 400 to 500 m/s extends from approximately 5 to 15 m in depth, underlain by a lower-velocity layer with $v_s$ of around 320 m/s. The low-velocity zone between 25 to 45 m in depth and from 0.55 to 0.70 km in distance appears to be a lateral continuation of this very layer. The velocity structure we infer is consistent with previous studies in the region that were derived from dispersion curve inversion (*29*, *44*) and SASW site characterization (*41*).

## Spatiotemporal Groundwater Dynamics

Fig. 2c shows the spatial and temporal distribution of traveltime delays of ballistic surface waves during the monitoring period. Time delays were computed through cross-correlations of (time-lapse) monitor and baseline data, followed by stacking at common mid-point locations. The baseline is defined as the average between July 7 and August 25, 2022 – the driest period of the study. Fig. 2d presents the averaged traveltime changes (blue line) across all spatial locations and daily precipitation (red line). Traveltime changes are observed to correlate with seasonal rainfall and drying cycles. The rainfall in wet seasons witnesses substantial traveltime reductions (blue regions), whereas lower traveltime delays occur during dry periods. Spatially, notable traveltime changes are concentrated between 0.1 and 0.45 km, whereas changes between 0.45 and 0.8 km are less pronounced. Seismic traces from grassy and paved locations further illustrate these contrasts (see Supplementary Fig. S2). Data beyond 0.8 km is increasingly noisy due to less satisfactory raw data quality.

Daily FWI results reveal temporal, depth-dependent, and lateral variations in seismic velocity changes (see Fig. 3 and Supplementary Fig. S6). Fig. 3a shows velocity changes as distance-time profiles averaged over depths of 15–30 m, while Fig. 3b presents distance-depth profiles at eight key moments (see red stars in Fig. 3a). Early in the monitoring period, on February 5, 2022, for example, seismic velocities were slightly lower than the baseline level, with the latter close to that on August 5, 2022. By February 2023, reduced velocity was observed following rainfall during the wet season. These negative changes intensified until April, reaching their peak on April 13, 2023, which coincided with the largest traveltime delays (see Fig. 2d). As drying progressed, decreased velocities began to recover toward the baseline level as shown by the image in June 2023. By August 2023, during continued dry conditions, the velocities were found to remain lower than those observed in August 2022, suggesting a net increase in groundwater storage over the past water year. Another wet season from November 2023 to March 2024 also witnessed negative velocity changes, although they were less pronounced than the previous year.

Spatially, we notice that velocity changes are mainly confined to depths below 15 m, with minimal variations detected in the shallower layers (see Fig. 3 and Fig. 4a). This depth corresponds to the approximate local water table depth that fluctuates between 13 and 22 m. The most significant changes are observed within the low-velocity layer at depths of 20 to 30 m (see Fig. 4a), which corresponds to the low-velocity layer shown the baseline model (Fig. 2b). In the shallow unsaturated zone (above the water table), the changes in water saturation on $v_s$ primarily caused changes in density but has limited influence on shear modulus (*20*). These minimal changes are not detected using surface wave inversion here. P-wave data can provide better delineation of saturation-driven medium changes in the unsaturated zone. A checkerboard resolution tests suggest that velocity changes to depths of at least 35 m can be resolved (see Supplementary Fig. S5).

Daily FWI also reveals pronounced lateral variability in seismic velocity changes. The most substantial velocity changes appear to occur beneath grassy areas (0.1 to 0.45 km in distance), whereas paved areas (0.45 to 0.65 km in distance) exhibit changes of smaller magnitude. These lateral variations are consistent with traveltime delay trends observed in Fig. 2c, while the time-lapse FWI analysis additionally capture these changes in $v_s$ model, a step further beyond only providing time series for characterizing the medium evolution. These results emphasize the influence of surface conditions and land use on the natural recharge of groundwater, where the grassy area has experienced greater groundwater infiltration recharge while paved areas register minute changes (*45*, *46*). Still, subsurface aquifer systems are interconnected, and velocity changes under paved areas may reflect water transport via preferential flow paths, as conceptualized in Fig. 4e. These findings suggest that urbanization, particularly the expansion of impervious surfaces, can reduce the aquifer recharge efficiency. Quantitative understanding of these interactions is crucial for planning for sustainable water resource management and building climate-resilient infrastructures.

## Physical Mechanisms of Velocity Changes

Near-surface seismic velocity variations can be attributed to multiple factors, including thermal-elastic stress induced by temperature changes and mechanical alterations driven by barometric pressure (*17*, *18*), and hydrological forces such as water saturation, clay hydration, and pore pressure changes (*20*, *47*, *48*). Additionally, the cooling effect of cold winter rainfall may lead to temperature-induced variations in seismic velocity (*49*).

With the time-lapse FWI images, we observe velocity changes predominantly in the saturated zone below the local groundwater table. These observed velocity changes are likely driven primarily by the shallow sediment's poroelastic response to precipitations. To quantify the poroelastic effect, we simulated the pore pressure changes at depth in response to precipitation at the ground surface using a basic model that couples fluid flow and mechanical response (see Materials and Methods: Simulation of pore pressure fluctuations). In Fig. 4b, we compare the simulated pore pressure (red line) and velocity changes at grassy areas averaged over a 200–400 m lateral range and a 20–30 m depth in the percentage of the baseline velocity model (blue line). In the saturated zone, rising water levels increase pore pressure ($P_{\text{pore}}$), which reduces the effective stress ($P_{\text{eff}}$), according to the relationship $P_{\text{eff}} = P_{\text{c}} - P_{\text{pore}} + S_W \psi$, where $P_{\text{c}}$ is the confining pressure (composed of hydrostatic and lithostatic stresses), $S_W$ is water saturation and $\psi$ is matric potential (*50*). Thereby, $v_s$ correlates positively with effective stress ($P_{\text{eff}}$) and correlates negatively with pore pressure (see Fig. 4e).

Fig. 4b shows a neat correlation between simulated pore pressure increase, derived with the poroelastic modeling, and the observed velocity decreases. During intense rainfall from late December 2022 to mid-January 2023, pore pressure increased sharply, coinciding with up to 1.7% velocity reductions. A similar pattern occurred from February to March 2023, with a velocity reduction of approximately 1.2%. In comparison, rainfalls in wet seasons of 2021 and 2023 also led to increased pore pressure and reduced velocities, but the effects were less pronounced due to the deficient amount of precipitation. The good agreement between simulated pore pressure increases and observed velocity reductions suggests that the poroelastic effect is likely the primary driving force of the detected velocity changes.

**Impact of Extreme Precipitation Events on Groundwater Levels**

The good alignment between FWI-derived velocity changes and in-situ groundwater table measurements (Fig. 4a) highlights the potential of time-lapse FWI as an effective tool for tracking daily water table fluctuation. While hydraulic head measurements are often spatially sparse and temporally discrete, daily FWI analysis can capture both transient and long-term aquifer dynamics. During the wet season in Water Year 2023, consecutive atmospheric river events caused significant recovery of aquifer storage, resulting in cumulative groundwater level rises of approximately 9 m (see Fig. 4a). These subsurface water storage changes corresponded to a 2.9% reduction in $v_s$, relative to the driest period between July to August 2022.

Notably, an EPE on December 31, 2022, led to a record-high daily precipitation of approximately 0.127 m. The abrupt rise in groundwater table height, referred to as water-table displacement ($\Delta\text{WTD}_{max}$), quantifies the increase in groundwater levels relative to pre-event conditions. Hydraulic head monthly measurements from the nearby well 06S03W02D032, however, failed to capture the $\Delta\text{WTD}_{max}$ for this EPE event due to its inadequate temporal sampling, as the hydraulic head was only measured on December 28, 2022, and January 31, 2023. By contrast, daily FWI analysis allows us to detect and quantify velocity changes in response to this EPE, revealing a rapid, substantial groundwater table rise within one single day (see Fig. 4d). In the following days, velocity changes were less pronounced due to less precipitation. Our analysis agrees with modeling studies that estimate $\Delta\mathrm{WTD}_{max}$ values between 0.6 and 2.4 m. Besides, the 2-year velocity change observations align with the understanding from previous case studies that a small number of large precipitation events could drive the majority of groundwater recharge (*51*, *52*).

Additionally, the recession time, which is defined as the period required for the water table to return to within 5% of $\Delta\mathrm{WTD}_{max}$, is estimated from daily FWI results to be more than 6 months (see Fig. 4a). As recession time is inversely related to hydraulic diffusivity (*53*), FWI-derived seismic velocity changes offer insights into key hydrological properties such as porosity and diffusivity. With additional in-situ data, calibrated rock physics models, and hydrologic models, our analysis could enable quantitative characterization of hydraulic processes, paralleling workflows used in hydrocarbon reservoir monitoring (*54*).

## Discussion

The innovative Vehicle-DAS approach introduced in this study advances the seismic monitoring and imaging of groundwater systems, offering high spatiotemporal resolutions that enables the localization of velocity changes at a daily basis, which is an important step beyond the conventional coda-wave monitoring methods. FWI resolves subsurface heterogeneous structures and changes at high resolution, which could better inform modeling of aquifer dynamics, recharge processes, and responses to hydrological forcing. Nonetheless, the resolved depth in this case study is limited by the absence of low-frequency data (<3 Hz) and the short offsets inherent to vehicle-induced sources. Additionally, thermal-elastic effects and soil moisture variability may complicate the interpretation of observed velocity changes. Applying time-lapse FWI to data generated by low-frequency traffic sources when available, such as heavy trucks and trains, can potentially extend monitoring depth with low-frequency data. Besides, piezometric measurements from hydraulic wells can be used to calibrate groundwater table estimations from FWI results.

This pilot application showcases that time-lapse FWI, combined with DAS sensing and traffic seismic sources, is a cost-effective and scalable approach for daily monitoring of near-surface aquifer dynamics at unprecedented temporal and spatial resolutions. The observed velocity changes show strong correlations with seasonal hydrological patterns, which are interpreted as resulting from pore pressure variations following precipitation events. Our results suggest a significant 2.9% $v_s$ reduction during the wet season of Water Year 2023. It shows that surface conditions, such as grassy versus paved areas, influence the effectiveness and efficiency of groundwater natural recharge from precipitations. The observed high spatiotemporal velocity changes enable the tracking of groundwater table, provide insights into the impacts of EPE on the subsurface, and offer the potential for informing aquifer properties such as hydraulic diffusivity. These findings highlight the promise of our approach to deliver meter-scale spatial resolution and daily temporal monitoring of subsurface hydrological changes, in support of more effective groundwater management.

# Materials and Methods

## Data description and pre-processing

We conducted daily groundwater monitoring along a 1 km section of Sandhill Road, California. The continuous DAS recordings were collected using a Luna OptaSense ODH-3 interrogator with a sampling rate of 250 Hz, a gauge length of 16 m, and a channel spacing of 8.16 m. To geo-locate each DAS channel accurately, we performed a tape test with a GPS-equipped vehicle (*28*, *55*).

The raw DAS data were collected in one-minute windows. For processing, we combined the data into 10-minute segments, applying a 5% taper to the ends of each segment to reduce edge effects. We then performed linear detrending and median removal to eliminate constant offsets. Fig. 1c demonstrates an example of the processed DAS data, including two primary signal types: quasi-static deformation signals caused by vehicle weight (frequencies below 1 Hz) and surface waves induced by vehicle movement (frequencies above 1 Hz) (*44*). The quasi-static signals were purposed for vehicle tracking, while the surface waves were utilized for targeted interferometry.

## Time-lapse targeted interferometry

To retrieve VSGs with a high signal-to-noise ratio (SNR), we applied the targeted interferometry workflow (*30*) to daily DAS data collected from December 2, 2021, to March 22, 2024, during the less congested period between 00:00 AM and 06:00 AM local time, when individual vehicles could be more easily separated. This workflow begins by precisely tracking vehicle trajectories in time and space using low-frequency quasi-static ground deformation signals (frequencies below 1 Hz) (*56*), as shown by the red lines in Fig. 1c. Once individual vehicle trajectories are identified, time-space windows containing coherent surface waves are selected (see orange boxes in Fig. 1c or enlarged view in Fig. 1d). These windows are constrained to a 22-second duration to minimize interference from other vehicles. Cross-correlation interferometry is then performed within these windows, focusing on surface wave energy (frequencies above 1 Hz) to retrieve the VSGs. Finally, VSGs from multiple vehicles within a given monitoring period (e.g., one day) are stacked to produce a time-lapse seismic survey (see Supplementary Fig. S1 for the daily number of detected vehicles). This workflow yields daily time-lapse seismic surveys with 39 virtual sources located from 146.88 m to 895.56 m at a 20 m interval, and 33 virtual receivers with a spacing of 8.16 m. An example VSG for a source located at 610 m along the fiber, stacked from 125 detected vehicles on December 3, 2022, is shown in Fig. 1e.

We define the baseline as the average of data between July 7 and August 25, 2022, representing the driest period of the monitoring interval. Also, a 14-day moving window average was applied to the daily data to enhance SNR further. The time-lapse seismic data were pre-processed using bandpass filtering (3–10 Hz), F-K filtering (200–750 m/s) to remove coherent noise, and trace-by-trace normalization to equalize amplitude variations. Traffic generates surface waves up to 25 Hz, but the 16 m gauge length acts as a spatial low-pass filter effect. Also considering the data quality, we chose frequencies up to 10 Hz for monitoring. Top mutes were applied to exclude anti-causal signals (very weak compared to the causal part). Two additional criteria were applied to ensure data quality. First, monitor traces with time delays exceeding 25 ms, as determined by cross-correlation with baseline traces, were identified as outliers and removed. Second, trace repeatability was assessed using the Normalized Root Mean Square (NRMS) (*57*), defined as:

$$\mathrm{NRMS}(a_t, b_t) = \frac{200\times\mathrm{RMS}(a_t - b_t)}{\mathrm{RMS}(a_t)+\mathrm{RMS}(b_t)}, \tag{1}$$

where RMS denotes the Root Mean Square. Spurious traces with NRMS values greater than 0.6 were muted, indicating poor repeatability and potential time-lapse noise in land seismic surveys (*58*).

Time-lapse seismic waveforms demonstrate sensitivity to subsurface changes and exhibit high repeatability (see Supplementary Fig. S2). Seismic traces recorded in a grassy area with a virtual receiver at 320 m reveal temporal variations in waveform arrivals, particularly during the winter of 2022 and spring of 2023, coinciding with intense storms. These variations are evident in the variation of time lags between baseline and monitored traces (red line in Supplementary Fig. S2). Elevated NRMS values (sky-blue line in Supplementary Fig. S2) during rainfall periods indicate substantial waveform changes and significant subsurface velocity perturbations. In contrast, time-lapse waveforms from a paved parking lot, with a virtual receiver at 520 m, show minimal temporal variations, possibly due to the impermeable surface limiting water infiltration. Correspondingly, time delays of monitoring data are negligible, and NRMS values consistently remain below 0.25, reflecting the high repeatability of time-lapse interferometry.

**Time-lapse elastic full-waveform inversion**

Time-lapse FWI is a high-resolution monitoring method that detects subtle variations in subsurface elastic properties by inverting for full seismic waveforms from repeated seismic surveys. Compared to conventional tomographic methods using traveltime information, time-lapse FWI provides significantly improved accuracy and resolution (*54*). In this study, we apply elastic time-lapse FWI to monitor subsurface changes associated with hydrological processes by inverting VSGs of Rayleigh surface waves recorded by dense DAS arrays and generated by vehicular sources.

Elastic wave propagation in the isotropic medium can be expressed by the stress-velocity formulation of the wave equation:

$$\begin{aligned} \rho\frac{\partial v_i}{\partial t} &= \frac{\partial \sigma_{ij}}{\partial x_j} + f_i \\ \frac{\partial \sigma_{ij}}{\partial t} &= \lambda\frac{\partial v_k}{\partial x_k}\delta_{ij} + \mu\left(\frac{\partial v_i}{\partial x_j} + \frac{\partial v_j}{\partial x_i}\right), \end{aligned} \tag{2}$$

with the free-surface boundary condition:

$$\sigma_{iz} = 0 \ \text{ at } \ z = 0, \tag{3}$$

where $\sigma_{ij}$ denotes the stress tensor, $v_i$ denotes the particle velocity in the $i$ direction, $x_i$ and $x_j$ denote the spatial coordinates, $\delta_{ij}$ denotes the Kronecker delta, and $f_i$ denotes the body force per unit volume. The elastic properties are described by Lamé parameters $\lambda$ and $\mu$, and the density $\rho$. As the inversion is driven by Rayleigh surface waves mainly sensitive to S-wave velocity $v_s = \sqrt{\mu/\rho}$, we only invert for $v_s$ and its temporal variations. P-wave velocity $v_p = \sqrt{(\lambda + 2\mu)/\rho}$ and density $\rho$ are empirically derived from $v_s$, with $v_p = 1.732 v_s$ and density $\rho$ following Gardner's relationship (*59*), with a chain rule taken into account.

For the baseline inversion, we employ the global-correlation objective function (*60*), which measures the phase similarity between observed and synthetic normalized waveforms. Accurate amplitude modeling for surface-wave FWI in land data is challenging due to near-surface scattering, attenuation, and imperfect fiber-ground coupling. The adopted global-correlation objective function is more tolerant of inaccuracies in modeled amplitudes and has been successfully applied in previous studies (*61*, *62*), which is defined as follows:

$$\phi_{bl} = \sum_{s=1}^{n_s} \sum_{r=1}^{n_r} \left[ -\hat{\mathbf{d}}_{s,r}^{\text{obs}} \cdot \hat{\mathbf{d}}_{s,r}^{\text{syn}} \right], \tag{4}$$

where $\hat{\mathbf{d}}_{s,r}^{\text{obs}}$ and $\hat{\mathbf{d}}_{s,r}^{\text{syn}}$ denote the trace-by-trace normalized observed and synthetic waveforms at the $s$-th source and $r$-th receiver, and the dot denotes the dot product of two vectors. The negative sign is added for optimization purposes. Synthetic data is modeled based on the principle of DAS sensing, which measures the averaged tangential strain over the gauge length at each channel. For the horizontally oriented fiber used in this study, the DAS data is modeled as follows (*63*):

$$d(x_r) = \frac{1}{gl} \left[ v_x \left( x_r + \frac{gl}{2} \right) - v_x \left( x_r - \frac{gl}{2} \right) \right], \tag{5}$$

where $gl$ denotes the gauge length and $d(x_r)$ denotes the DAS data at channel $x_r$.

For the monitor FWI, we use the double-difference (DD) inversion strategy (*64*, *65*) to invert the differences between baseline and monitor waveforms. This approach minimizes the impact of any inaccuracies in the baseline model, thereby reducing potential inversion artifacts. However, the conventional DD strategy based on the L2 norm is sensitive to errors in both absolute amplitude and phase. To address these limitations, we modify the traditional DD approach by using the global-correlation misfit function as in the baseline FWI. Thus, the objective function $\phi_{ml}$ for the monitor inversion is defined as:

$$\phi_{ml} = \sum_{s=1}^{n_s} \sum_{r=1}^{n_r} \left[ -\hat{\mathbf{d}}_{s,r}^{\text{com}} \cdot \hat{\mathbf{d}}_{s,r}^{\text{syn}} \right], \tag{6}$$

where $\hat{\mathbf{d}}_{s,r}^{\text{syn}}$ denotes the normalized synthetic data computed from the monitor model, and $\hat{\mathbf{d}}_{s,r}^{\text{com}}$ represents the composite data defined as follows:

$$\hat{\mathbf{d}}^{\text{com}} = \hat{\mathbf{d}}_{\text{ml}}^{\text{obs}} - \hat{\mathbf{d}}_{\text{bl}}^{\text{obs}} + \hat{\mathbf{d}}_{\text{bl}}^{\text{syn}}, \tag{7}$$

where $\hat{\mathbf{d}}_{\mathrm{bl}}^{\mathrm{syn}}$ denotes the normalized synthetic data calculated from the baseline model, and $\hat{\mathbf{d}}_{\mathrm{ml}}^{\mathrm{obs}}$ and $\hat{\mathbf{d}}_{\mathrm{bl}}^{\mathrm{obs}}$ represent the observed monitor and baseline data, respectively.

Numerical simulations are performed using the staggered grid finite-difference method (*66*, *67*), with fourth-order accuracy in space and second-order accuracy in time. The W-AFDA scheme (*68*) is used to model the free-surface conditions for Rayleigh surface waves, while other boundaries are set to be absorbing to eliminate boundary reflections. The adjoint-state method (*69*) is employed to compute the gradient of the model parameters for the wave equation operator. In DAS modeling, the data restriction operator acts as a first-order differential operator on the horizontal particle-velocity wavefield, and its adjoint operator is applied in the inversion. The *l*-BFGS optimization algorithm (*70*) is used to solve both the baseline and monitor FWI, with the model parameterization and objective function detailed above.

We perform the baseline FWI using a 1-D $v_s$ model (see Supplementary Fig. S3a). The source wavelet is extracted from the pre-processed VSGs by aligning and stacking ballistic surface waves and set to be consistent for all VSGs. The data in frequency bands of 3–6 Hz, 3–8 Hz, and 3–10 Hz are sequentially inverted. Our baseline FWI workflow resulted in kinematically improved waveform fitting for sources spanning the survey line (see Supplementary Fig. S4). Using the derived baseline model, we perform the time-lapse FWI to daily VSGs in the 3–10 Hz frequency range (see Supplementary Fig. S7 for the misfit reduction for daily FWI). The resolving ability is further validated through checkboard tests (see Supplementary Fig. S5).

**Simulation of pore pressure fluctuations**

To evaluate the poroelastic response of the shallow sediments to precipitations, we adopt the mechanical formulation in Roeloffs, 1988 (*71*), to calculate pore pressure changes at depth in a half-space corresponding to a step load at the surface as:

$$p(z,t) = \alpha p_0 \operatorname{erf}\left(\frac{z}{\sqrt{4ct}}\right) + p_0 \operatorname{erfc}\left(\frac{z}{\sqrt{4ct}}\right), \quad (8)$$

where $p_0$ denotes the amplitude of the surface step load, $z$ denotes the depth, $t$ denotes the time since the load being applied, $c$ denotes the hydraulic diffusivity, $\alpha$ denotes an elastic constant determined by the Skempton's coefficient $B$ and the undrained Poisson's ratio $v_u$ as $\alpha = \frac{B(1+v_u)}{3(1-v_u)}$, and erf and erfc are the error and the complementary error functions. This formulation is derived from a simplified model that takes into account the coupling of fluid flow and elasticity in saturated porous medium (*71*, *72*). The two terms in Eq. (8) together reflect the undrained and drained responses: The first term describes the undrained response that is an instantaneous change due to applied elastic loading and the poroelastic relaxation due to diffusion outside the model, and the second term describes the drained response due to the diffusion of fluid from the surface to depth (*71*, *73*).

For a series of precipitations, the surface load $\delta p_i$ on each day $i$ ($i = 1, 2 \ldots, n$) can be considered as the deviation of the precipitation rate from its average rate (*73*). The pore pressure changes at depth in response to the precipitation series can then be derived as:

$$P(z, t_i) = \frac{B(1+\upsilon_u)}{3(1-\upsilon_u)} \sum_{i=1}^{n} \delta p_i \mathrm{erf}\left(\frac{z}{\sqrt{4c(n-i)\delta t}}\right) + \sum_{i=1}^{n} \delta p_i \, \mathrm{erfc}\left(\frac{z}{\sqrt{4c(n-i)\delta t}}\right). \quad (9)$$

In this study, we use the precipitation record from a nearby meteorological station (KCAMENLO88, denoted in Fig. 1a) as the input of the pore pressure simulation with the hydraulic diffusivity equal to 0.1, the Skempton's coefficient equal to 0.8, and an undrained Poisson's ratio of 0.4. The annual cumulative precipitation and the modeled pore pressure variations are shown in Fig. 4c and b, respectively, in comparison with the observed velocity changes.

## Acknowledgments

We thank the affiliate companies of the Stanford Earth imaging Project (SEP) for their financial support. We acknowledge the Stanford ITS fiber team for their assistance during the Stanford DAS-2 experiment and the Stanford School of Earth IT team for hosting the DAS interrogator in the school's computer room. We also thank Luna-OptaSense for providing the DAS interrogator. HL acknowledges the support of Shirley A. & Stanley H. Ward Scholarship from the SEG Foundation.

## Author contributions:

Conceptualization: H.L., J.L., S.M., S.Y., R.C., and B.B.
Methodology: H.L., J.L., S.Y., and S.M.
Investigation: H.L., J.L., S.M., and R.C.
Data Collection: S.Y., R.C., B.B.
Visualization: H.L.
Supervision: B.B.
Writing—original draft: H.L.
Writing—review & editing: H.L., J.L., S.M., S.Y., R.C., and B.B.

**Competing interests:** The authors declare no competing interests.

## Data and materials availability:

The targeted interferometry code for virtual source gather retrieval is available at https://github.com/syyuan93/das_veh. The time-lapse FWI workflow is performed based on the deepwave package (https://github.com/ar4/deepwave). Daily FWI models and the analysis code can be accessed at https://doi.org/10.5281/zenodo.14597601, which also includes continuous DAS recordings and retrieved virtual source gathers from December 2022. Precipitation data are retrieved from https://www.wunderground.com/, and groundwater measurements are available at https://gis.valleywater.org/Wells.html.

## Figures

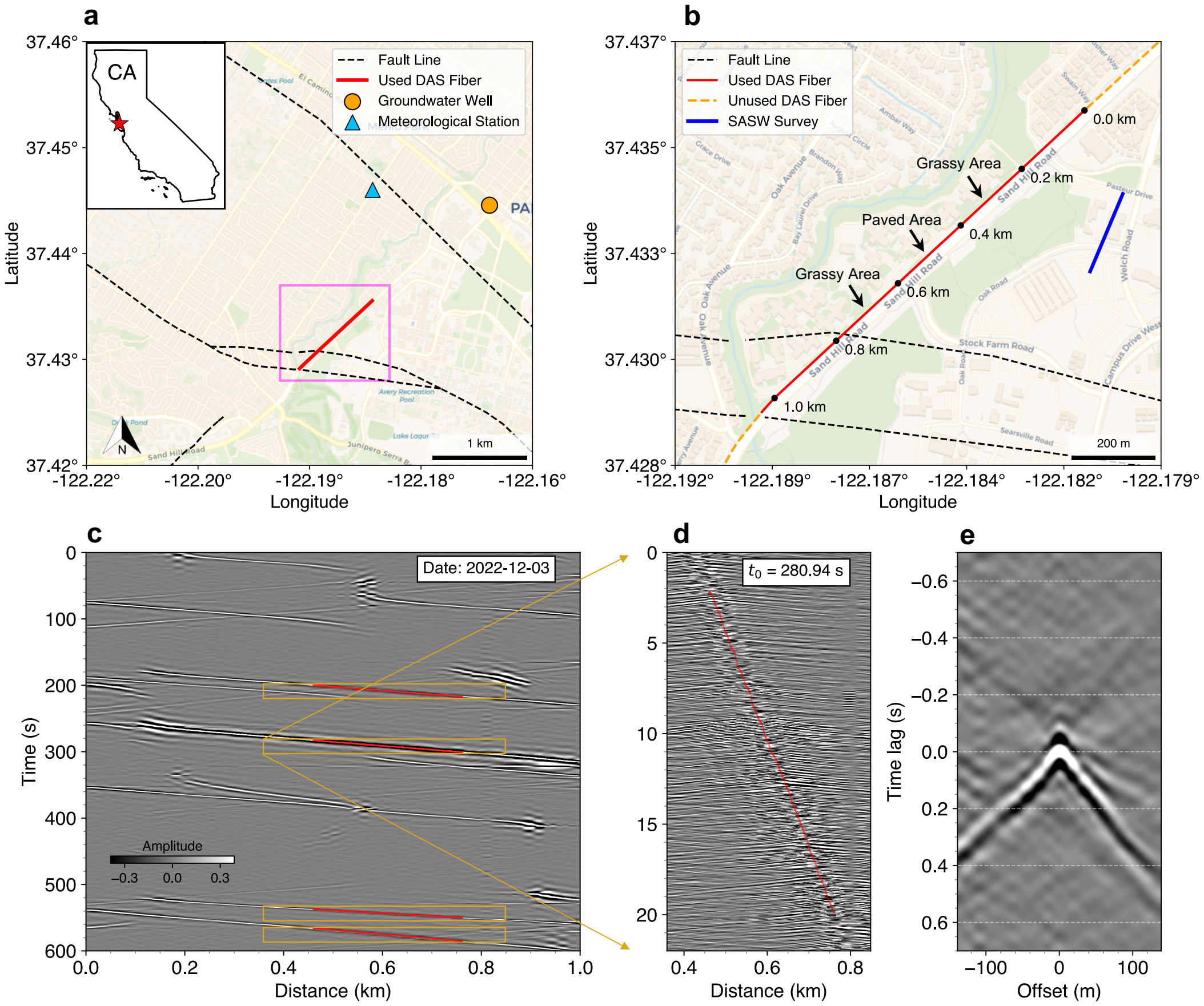

**Fig. 1. Targeted interferometry of vehicle-induced surface waves on Stanford DAS-2 array. a.** Map view of the study area and the used section of Stanford DAS-2 array (red line), as indicated by a red star in the inset California (CA) map. The orange circle marks the groundwater monitoring well (06S03W02D032), the blue triangle indicates the meteorological station (KCAMENLO88), and the dashed black lines represent the surface traces of local faults (Source: California Geological Survey, Quaternary fault and fold database for the United States, accessed December 1, 2024, at: https://www.usgs.gov/natural-hazards/earthquake-hazards/faults). **b.** The zoom-in view (purple box in panel **a**). The red line represents the fiber section used for groundwater monitoring, the dashed orange line indicates unused fiber sections, and the blue line marks the location of the Spectral Analysis of Surface Waves (SASW) survey. **c.** Example of preprocessed DAS data from a 10-minute recording on December 3, 2022, starting at 01:40:00 AM (local time). Red lines indicate tracked vehicle trajectories, and the orange boxes highlight selected time-space windows with induced surface waves. **d.** Example of vehicle-induced surface waves (bandpass filtered at 1–30 Hz) within a selected time-space window in panel **c**. **e.** Example of a virtual source gather (bandpass filtered at 3–10 Hz) at a source location of 610 m obtained using the targeted interferometry for December 3, 2022, by stacking individually retrieved VSG from 125 vehicles.

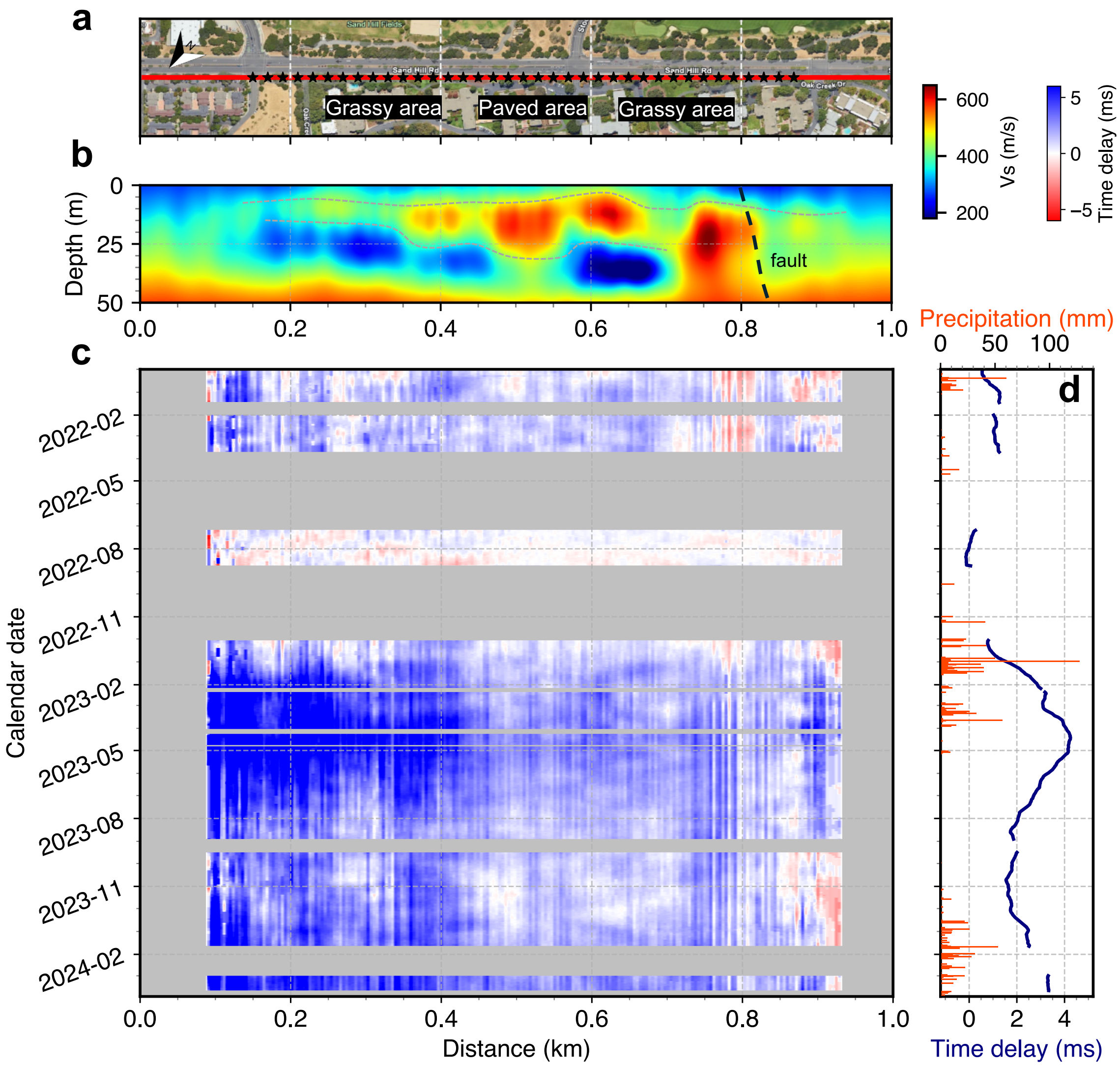

**Fig. 2. Near-surface structure variability and spatiotemporal monitoring of traveltime changes. a.** Map of the local landscape showing the DAS fiber (red line) and virtual source locations (black stars). **b.** Baseline FWI result ($v_s$) obtained using a multi-scale inversion. **c.** Maps of traveltime changes for data in the frequency of 3-10 Hz. These maps are obtained by first computing cross-correlation between monitor and baseline traces and then stacking at common mid-point locations. The baseline is defined as the average of data between July 7 and August 25, 2022. Positive time delays (blue) indicate delayed arrivals compared to the baseline, suggesting decreased subsurface velocity. Grey regions indicate areas lacking data from data gaps or common mid-point stacking. **d.** The stacked traveltime delay across all spatial locations (deep blue line) and recorded precipitation at meteorological station KCAMENLO88 (red line).

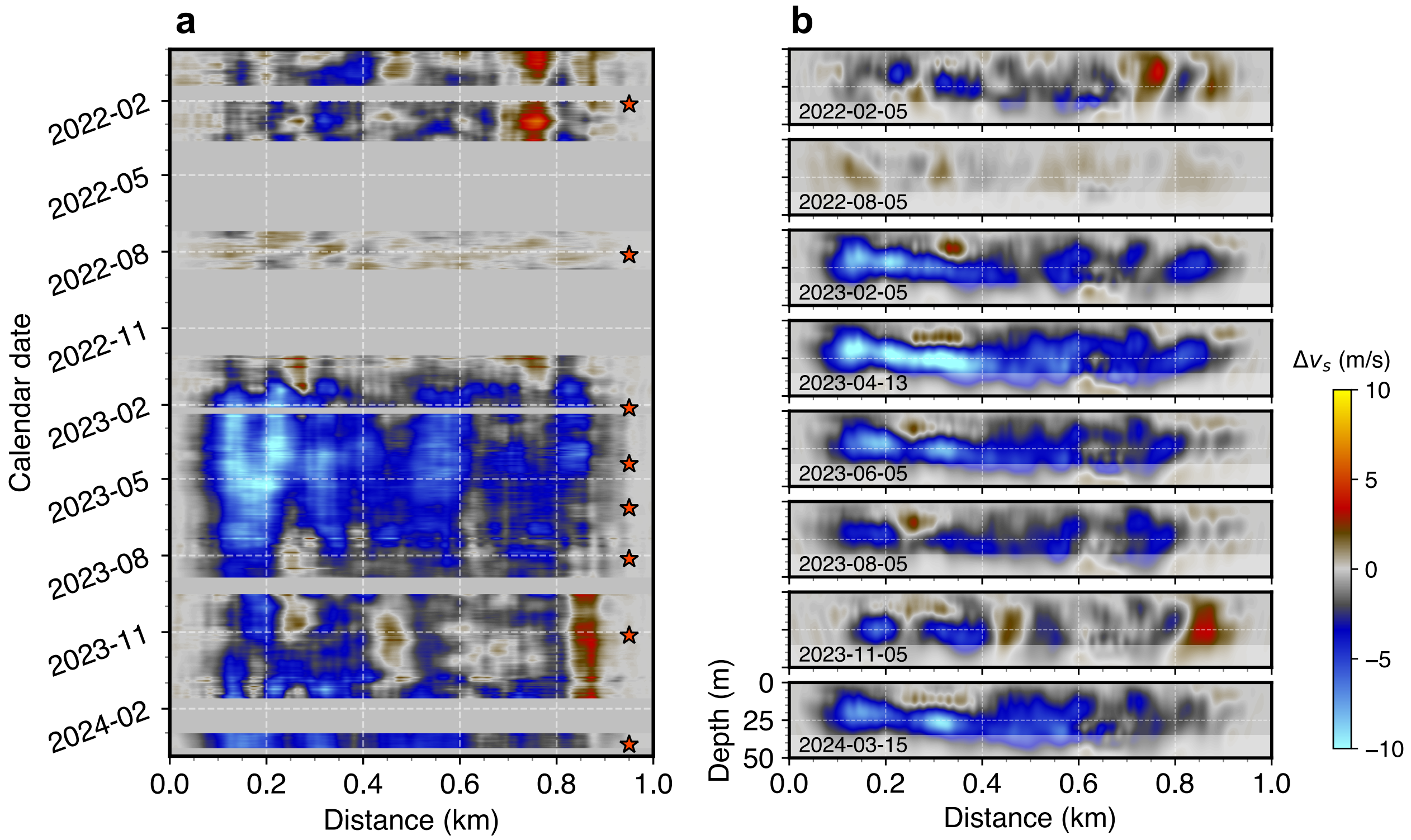

**Fig. 3. Spatiotemporal monitoring of subsurface velocity changes**. **a.** Time-lapse velocity changes in $v_s$ averaged between the depth range of 15–30 m. **b.** Distance-depth profiles for selected dates marked by stars in panel **a**. Particularly, the date of 2022-08-05 represents the driest period of the monitoring interval and 2023-04-13 the wettest period that corresponds to the largest time delay in Fig. 2d. The time-lapse FWI is performed with 3–10 Hz data. The shaded areas indicate the depth ranges with limited resolution.

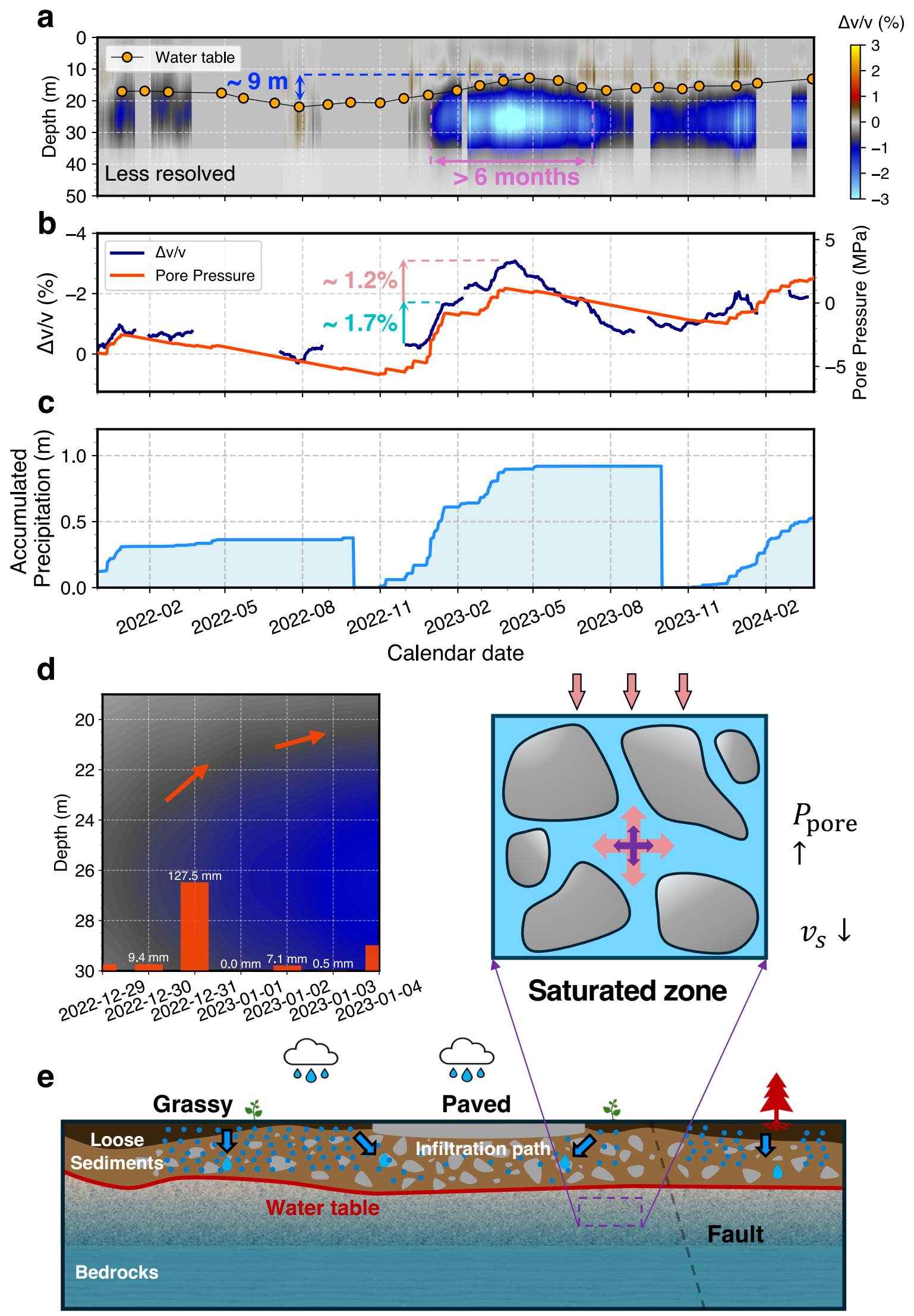

**Fig. 4. Interpretation of observed velocity changes and conceptual illustration of the near-surface groundwater system**. **a**. Depth-time profile of observed changes in $v_s$, averaged over the 200–400 m distance range. Orange dots represent groundwater table depth from a nearby monitoring well (06S03W02D032). The shaded areas indicate the less resolved depth ranges. **b.** Changes of $v_s$ in percentage relative to the baseline model (deep blue line), averaged over the 200–400 m distance range and a depth range of 20–30 m. Simulated pore pressure variations at a depth of 25 m (orange line) using precipitation records. **c.** Accumulated precipitation during the water year cycle, as recorded at meteorological station KCAMENLO88. **d.** Observed velocity changes resulting from the EPE on December 31, 2022, suggest a rapid rise in the groundwater table in one single day. **e.** Conceptual illustration portraying the near-surface groundwater system. In the saturated zone, pore pressure variations primarily drive velocity changes; increased pore pressure reduces $v_s$.

# Supplementary Materials

Figs. S1 to S7.

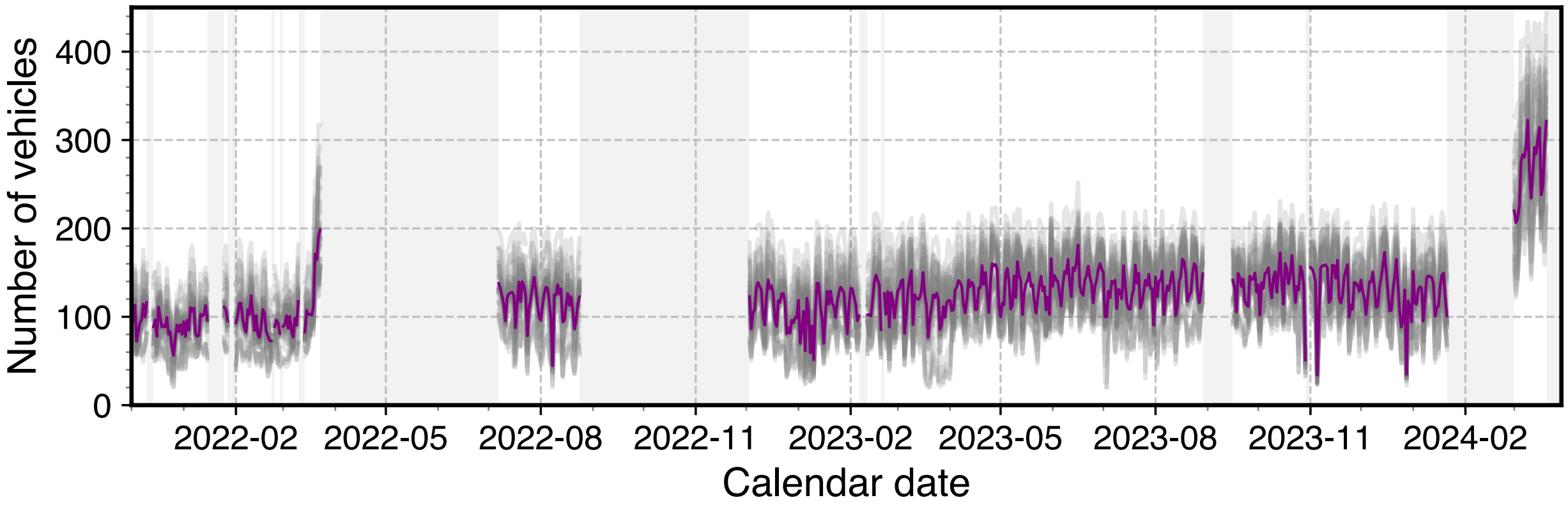

**Fig. S1.** Data availability and vehicle count for stacking daily VSGs. Number of vehicles detected and used for stacking VSGs across all sources during the monitoring period from December 1, 2021, to March 29, 2024. Different lines represent different source locations. The purple line denotes the average number of detected vehicles. The gray areas denote data gaps due to logistical interruptions.

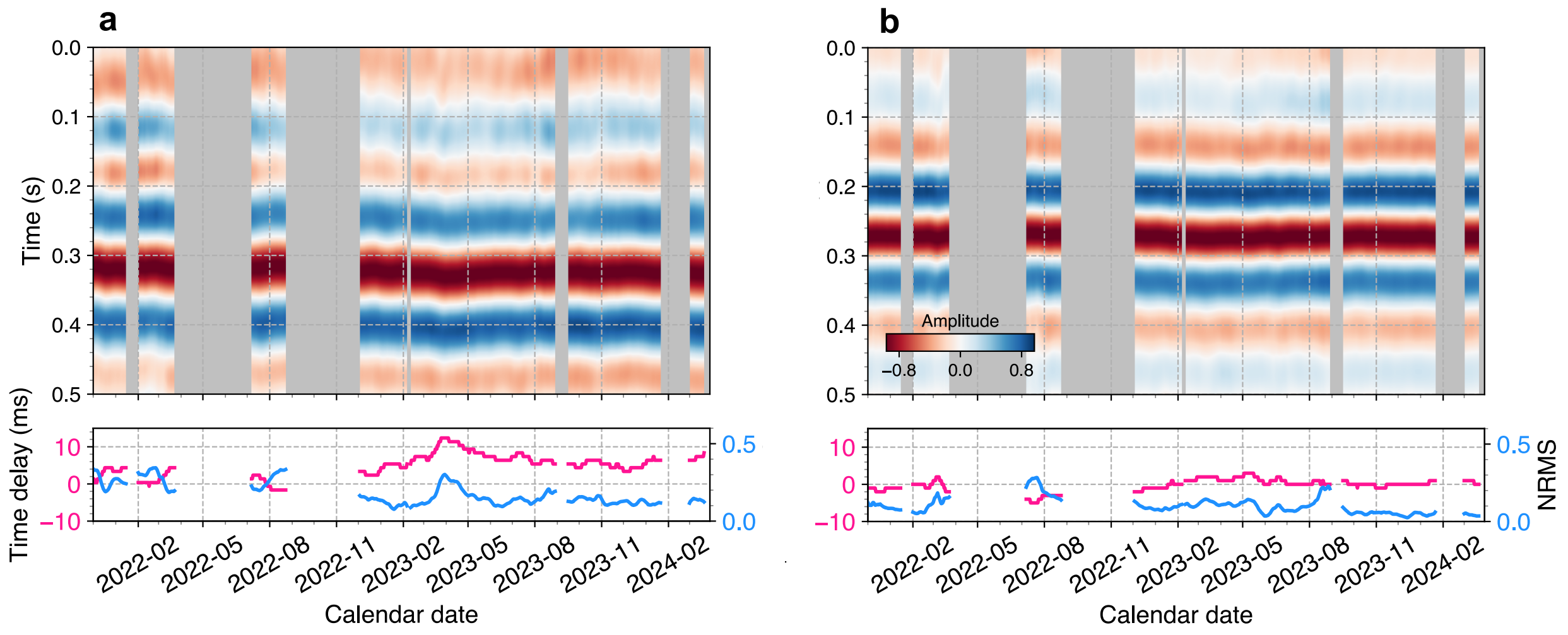

**Fig. S2. Time-lapse seismic waveforms under varying surface conditions. a.** Time-lapse trace at the location of 320 m (a grassy area). The lower subplot shows the time delay of the monitor trace relative to the baseline trace (red line) and the Normalized Root Mean Square (NRMS) value (blue line). **b.** Same as (a), but for a virtual source at 520 m (a paved area). Both traces have an offset of -90 m and are filtered within the 3.0 to 10.0 Hz frequency band. The baseline is defined as the average of data collected between July 7 and August 25, 2022, representing the driest period of this study. Gray regions mark periods with data gaps in the DAS recordings.

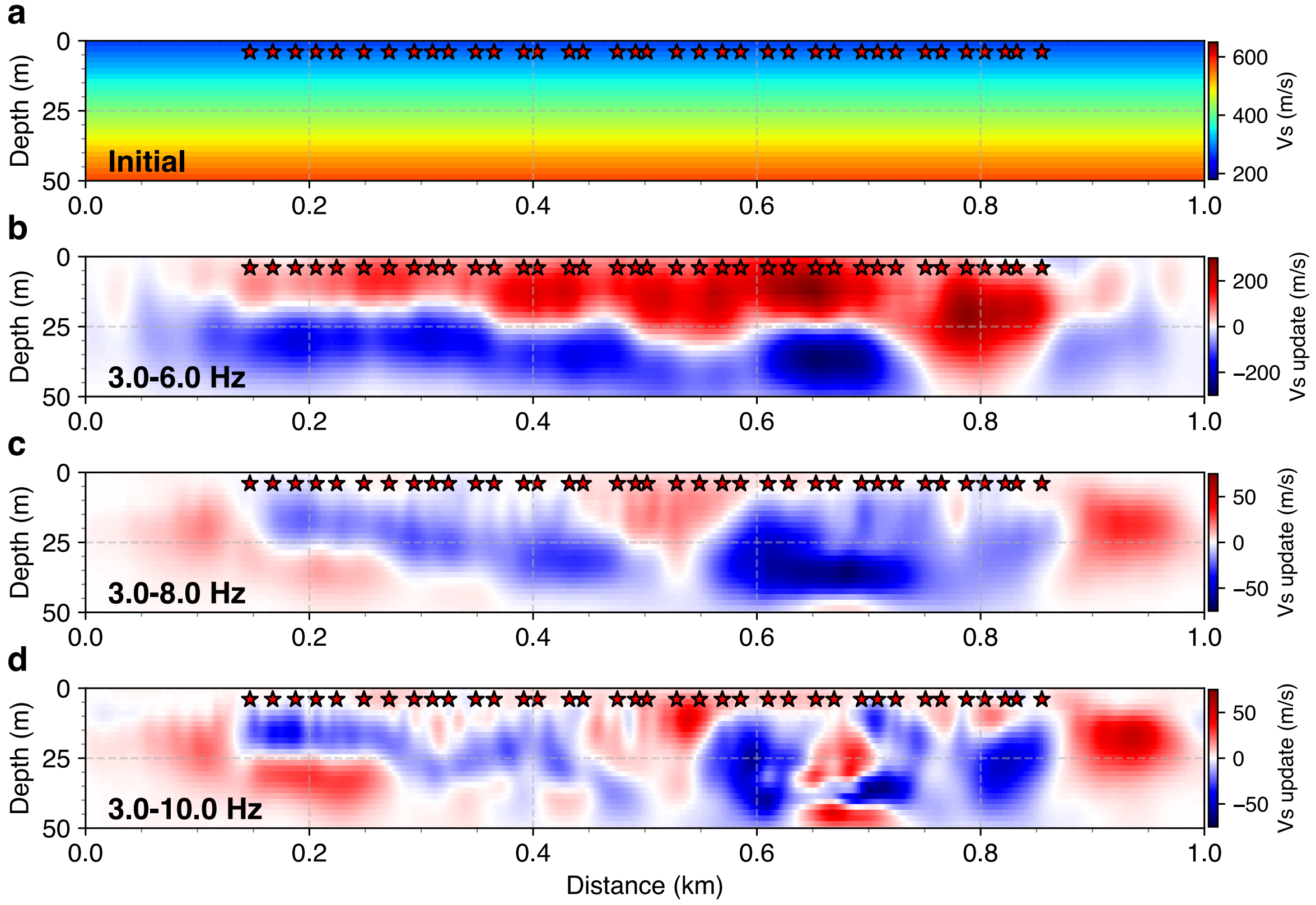

**Fig. S3. Initial velocity model and velocity updates in the multi-scale inversion. a.** The initial S-wave model ($\boldsymbol{v_s}$) used for the multi-scale FWI with velocities ranging from 270 m/s at the surface to 560 m/s at a depth of 50 m. **b.** FWI update in $\boldsymbol{v_s}$ in the 3-6 Hz frequency band, starting from the 1D initial model shown in panel **a**. **c.** FWI update in $\boldsymbol{v_s}$ in the 3-8 Hz frequency band, using the final model from the 3-6 Hz inversion as the starting model. **d.** FWI update in $\boldsymbol{v_s}$ in the 3-10 Hz frequency band, using the final model from the 3-8 Hz inversion as the starting model.

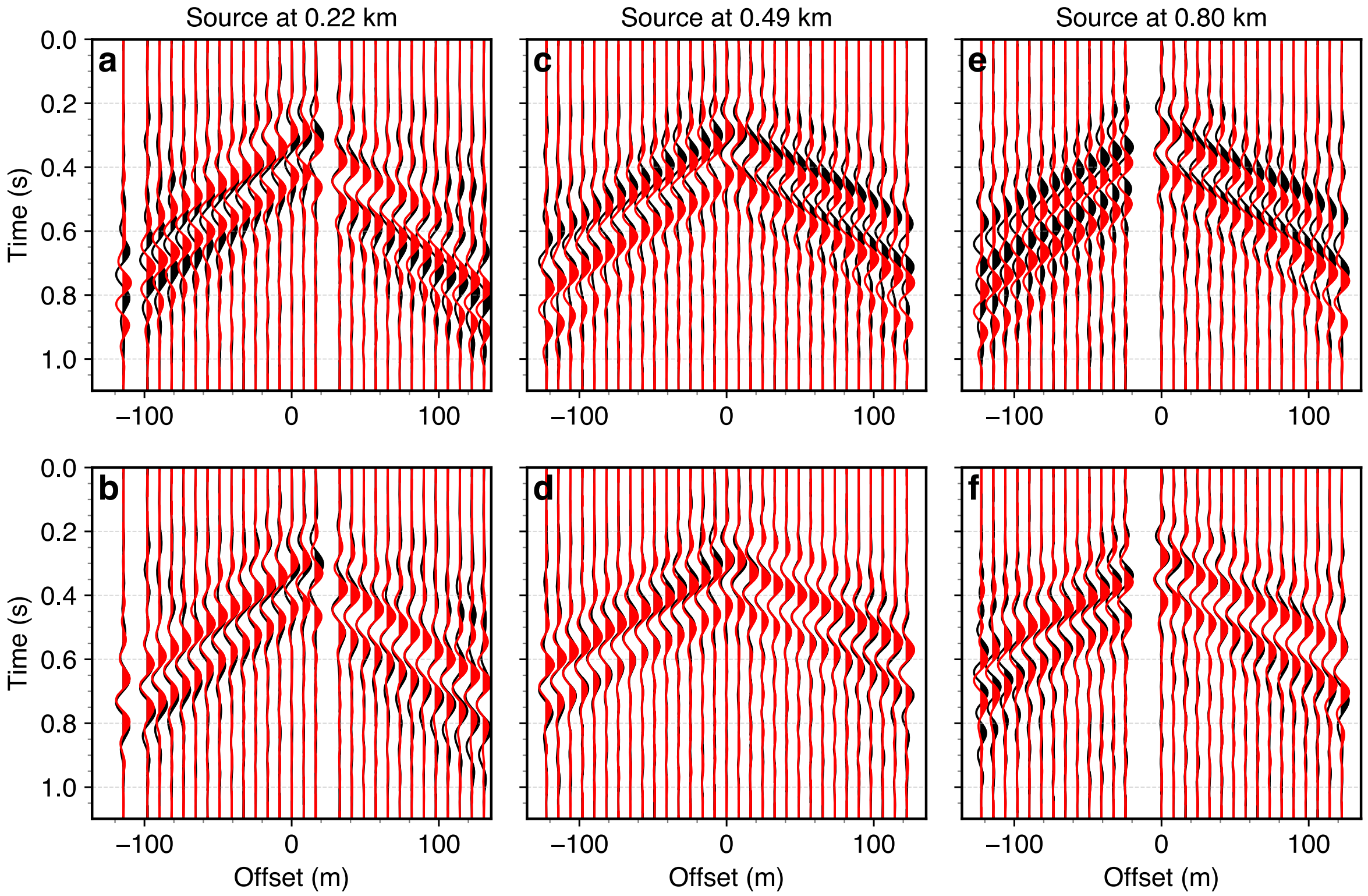

**Fig. S4. Waveform fitting before and after baseline FWI. a.** Waveform comparison for the source at 0.22 km, with observed data shown in black and synthetic waveforms from the 1D initial model (Fig. S2a) shown in red. **b.** Waveform comparison for the source at 0.22 km, with observed data shown in black and synthetic waveforms from the final FWI model (Fig. 2b in the main text) shown in red. **c.** and **d.** Same as panels **a** and **b** but for a source at 0.49 km, located in the middle of the survey line. **e.** and **f.** Same as panels **a** and **b** but for a source at 0.80 km, located at the end of the survey line. The data presented correspond to the final frequency band of 3–10 Hz in the multi-scale inversion. Good fitting is indicated by the good overlap between the red synthetic waveforms and the black observed waveforms.

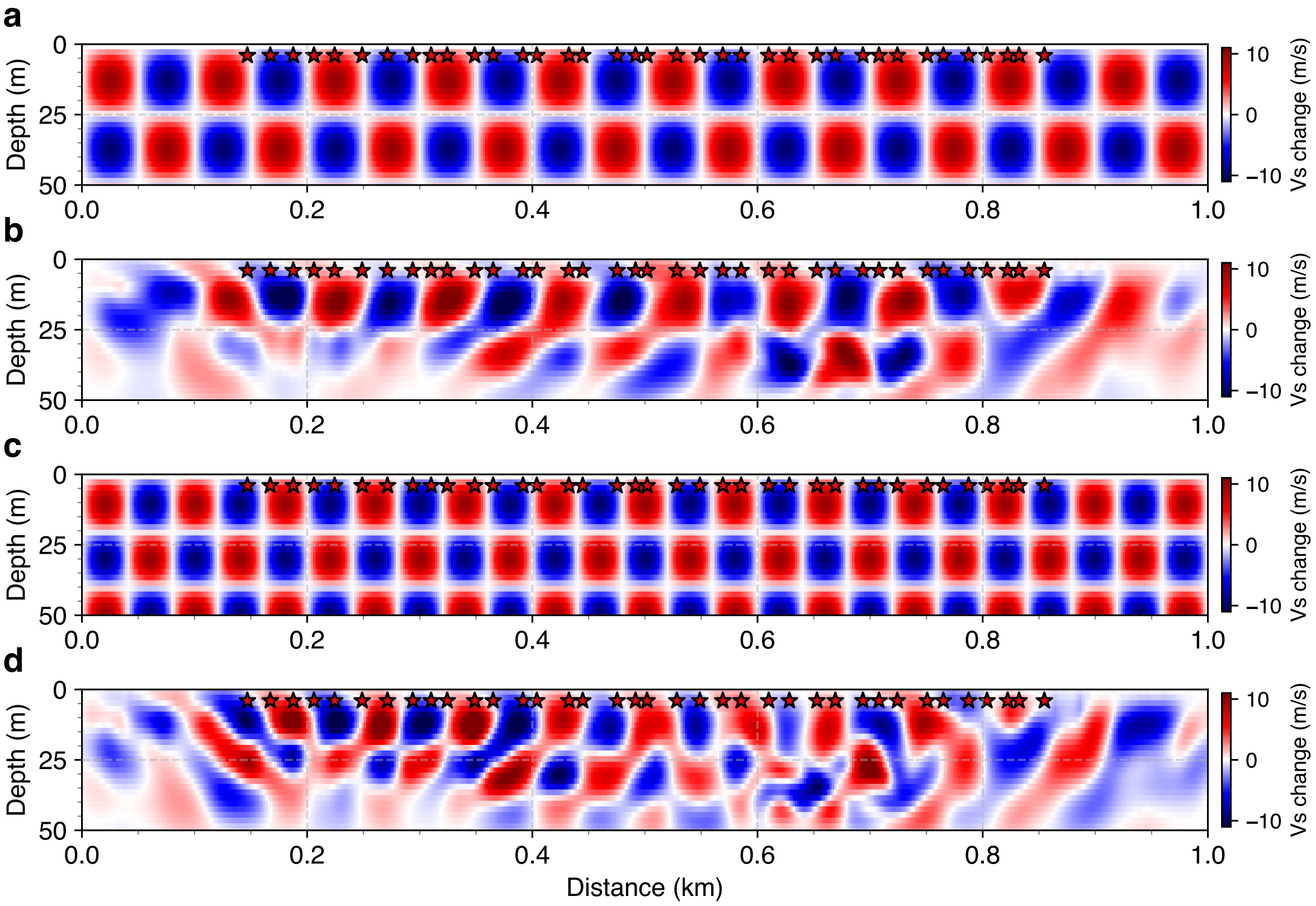

**Fig. S5. Checkerboard tests in time-lapse FWI for resolution analysis. a.** Checkerboard test with large-scale velocity anomalies measuring 25 m in depth and 50 m laterally, with a peak value of 10 m/s. These velocity anomalies are added to the FWI baseline model (Fig. 2b in the main text) to generate the *true* model to be recovered by time-lapse FWI. **b.** Inverted anomalies from the time-lapse FWI in the 3–10 Hz frequency band. The inversion parameters, including smoothness constraints, optimization techniques, acquisition setup, and bad trace muting, are identical to those used in the field-data FWI. The targeted anomalies are shown in panel **a**. **c.** Checkerboard test with smaller velocity anomalies measuring 20 m in depth and 40 m laterally, with a peak value of 10 m/s. **d.** Inverted anomalies from the time-lapse FWI in the 3–10 Hz frequency band, with *true* anomalies shown in panel **c**.

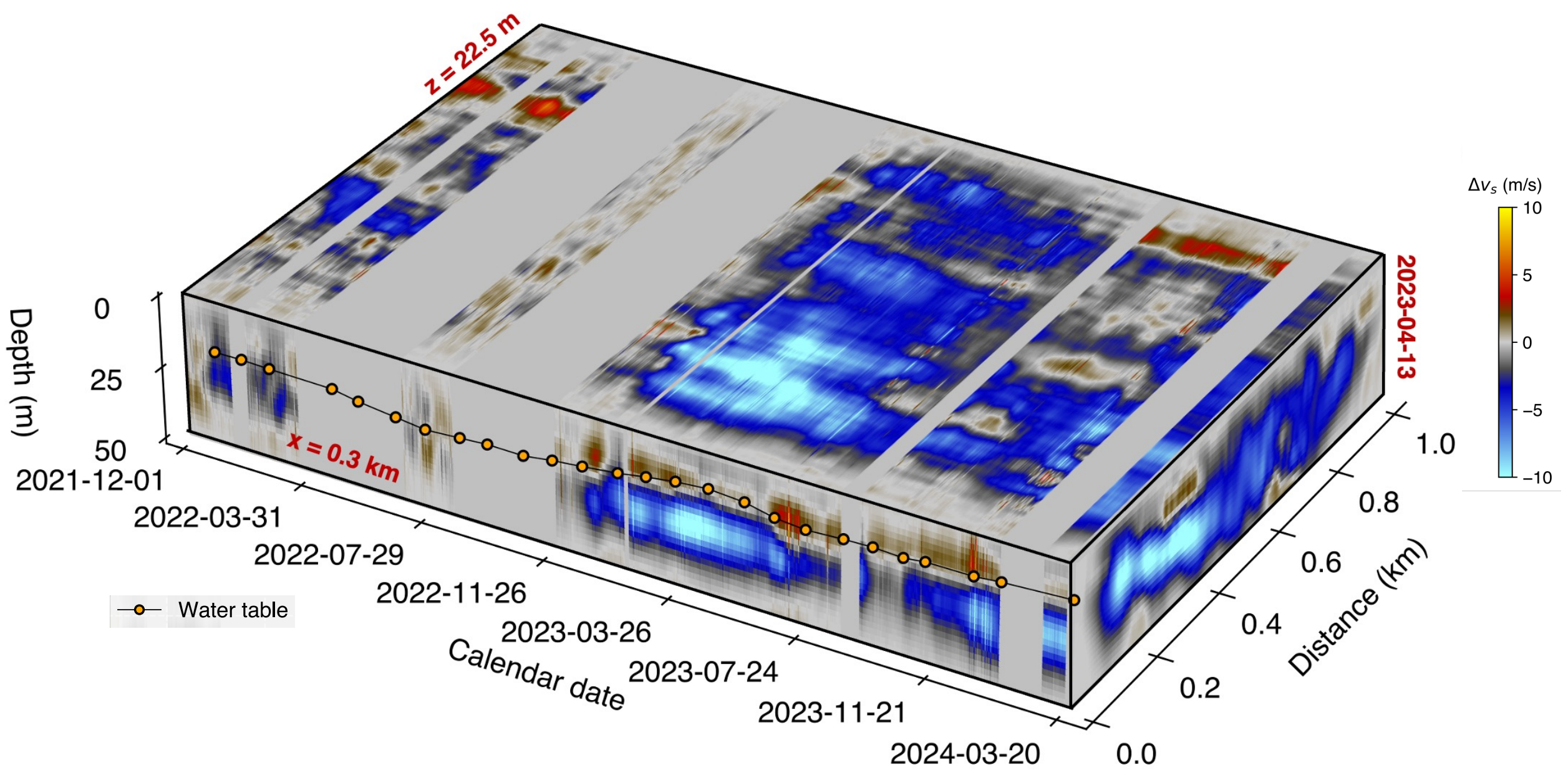

**Fig. S6. 3-D cube visualization of spatiotemporal subsurface changes in $v_s$.** The front view displays the time-depth profile at a distance of 0.3 km, the top view shows the time-distance profile at a depth of 22.5 m, and the right view presents the distance-depth profile for April 13, 2023. Orange dots show the in-situ groundwater table measurements from a nearby monitoring well (06S03W02D032). This figure extends the results shown in Figure 2 of the main text, providing an alternative visualization to highlight subsurface variations detected by daily FWI over time and space.

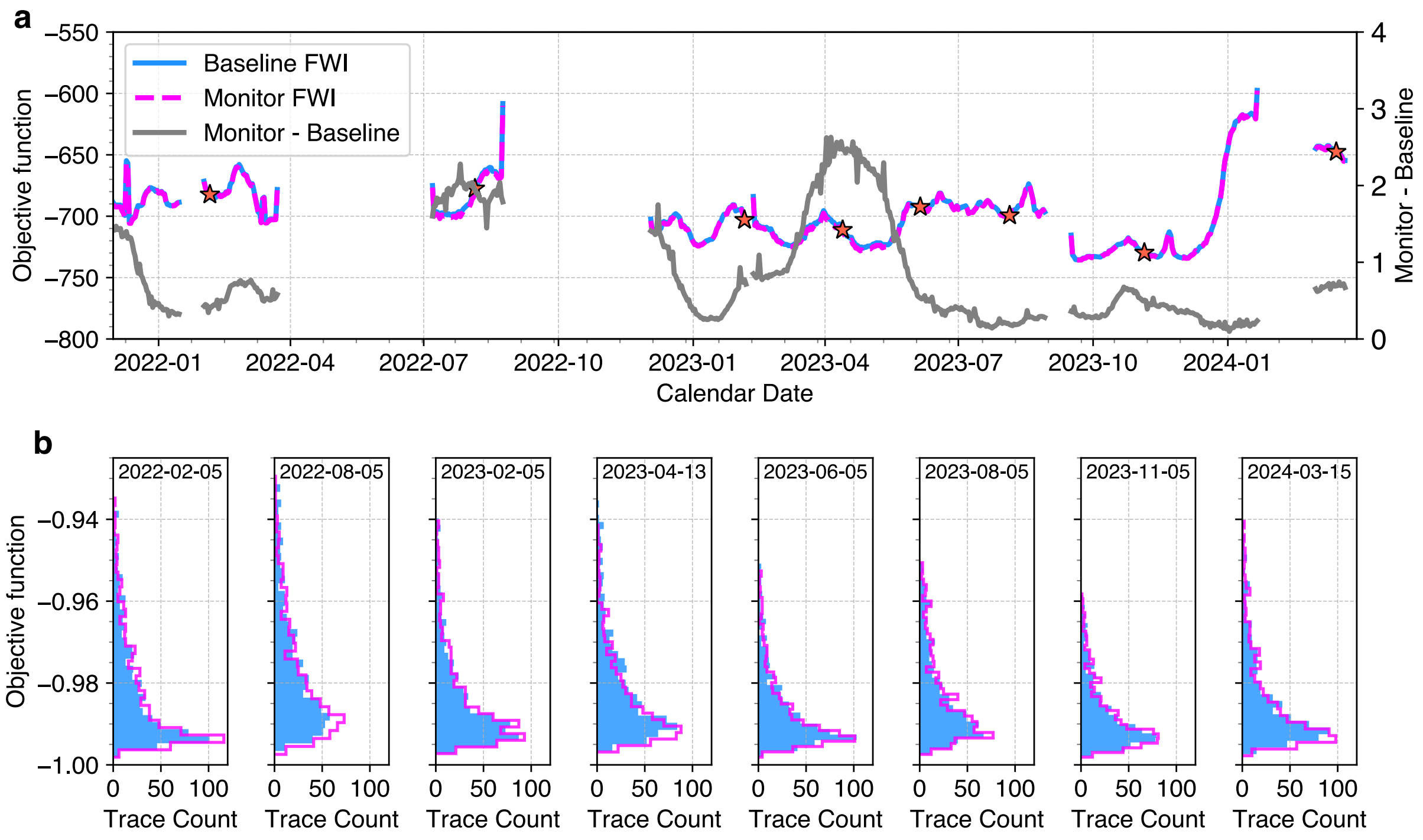

**Fig. S7. Misfit reduction in time-lapse FWI. a.** Total misfit for before monitor FWI (light blue) and after monitor FWI (magenta), summed over all traces included in the inversion. The gray line represents the difference between the baseline and monitor FWI misfits, illustrating the degree of misfit reduction achieved for each monitoring date. **b.** Histogram of the misfit for each trace used in the inversion. The light blue bars correspond to the misfit before the time-lapse FWI baseline FWI, and the magenta bars represent that after monitoring FWI. The selected dates shown here correspond to the monitor FWI results presented in Fig. 3b in the main text. Different dates exhibit varying degrees of misfit reduction, due to different levels of changes in subsurface velocity across the monitoring period.